\documentclass[aps,showpacs,twocolumn]{revtex4}
\usepackage[utf8x]{inputenc}
\usepackage{graphicx,subfigure,color}
\usepackage{latexsym,amsmath,amssymb,amsfonts,amsthm,enumerate}
\usepackage{amscd}
\usepackage{bm}
\usepackage{wasysym}
\usepackage{pict2e}
\usepackage{float}

\begin{document}
\bibliographystyle{apsrev}

\title{Majorization relations and entanglement generation in a beam splitter}

\author{C. N. Gagatsos}
\author{O. Oreshkov}
\author{N. J. Cerf}
\affiliation{Quantum Information and Communication, Ecole polytechnique de Bruxelles,
Université libre de Bruxelles, 1050 Brussels, Belgium}

\date{\today}

\begin{abstract}
We prove that a beam splitter, one of the most common optical components, fulfills several classes of majorization relations, which govern the amount of quantum entanglement that it can generate. First, we show that the state resulting from $k$ photons impinging on a beam splitter majorizes the corresponding state with any larger photon number $k'>k$, implying that the entanglement monotonically grows with $k$. Then, we examine parametric infinitesimal majorization relations as a function of the beam-splitter transmittance, and find that there exists a parameter region where majorization is again fulfilled, implying a monotonic increase of entanglement by moving towards a balanced beam splitter. We also identify regions with a majorization default, where the output states become incomparable. In this latter situation, we find examples where catalysis may nevertheless be used in order to recover majorization. The catalyst states can be as simple as a path-entangled single-photon state or a two-mode vacuum
squeezed state.
\end{abstract}

\maketitle

\section{Introduction}

In quantum optics, one of the most common transformations consists in the linear coupling between two modes of the electromagnetic field, as effected, for instance, by a beam splitter in bulk optics or an optical coupler in fiber optics \cite{book}. Mathematically, it corresponds to a rotation in phase space, namely
\begin{eqnarray}
\begin{pmatrix}
{\hat a}\\
{\hat b}
 \end{pmatrix}
 \to
\begin{pmatrix}
{\hat a}' \\
{\hat b}'
 \end{pmatrix}
=
\begin{pmatrix}
\cos \theta  & -\sin \theta\\
\sin \theta & \cos \theta
 \end{pmatrix}
 \begin{pmatrix}
{\hat a}\\
{\hat b}
 \end{pmatrix},
 \label{BS}
\end{eqnarray}
where ${\hat a}$ and ${\hat b}$ are the annihilation operators of the two modes that are coupled, while the angle $\theta \in [0, \pi/2]$ is a coupling parameter related to the transmittance $\tau=\cos^2 \theta$ of the beam splitter. The beam splitter is called balanced when $\tau=1/2$ or $\theta = \pi/4$. The transformation (\ref{BS}) belongs to the set of Gaussian unitaries as it corresponds, in the Heisenberg picture, to a linear canonical transformation in the annihilation (creation) operators ${\hat a}^{(\dagger)}$ and ${\hat b}^{(\dagger)}$, or equivalently to a quadratic Hamiltonian, namely  $H=i({\hat a}^{\dagger}{\hat b} - {\hat a}{\hat b}^{\dagger})$, see \cite{Gaussian} for a review on Gaussian transformations. Moreover, it is a passive linear operation since it conserves the total photon number, so it can be realized with a simple glass plate with a thin coating and no additional (pump) energy.

The beam splitter is very conveniently modeled within the so-called symplectic formalism by focusing on the action of the rotation (\ref{BS}) on the first- and second-order moments of the quadrature operators. This enables treating complex optical circuits made of beam splitters and other optical devices in a very concise way, which is sufficient for many purposes, e.g., when the goal is to predict distributions in phase space such as Wigner functions. However, if we want to make predictions about entropies or entanglement, we need to move back to state space and work with density operators. Such calculations are often nontrivial despite the simplicity of the transformation in phase-space representation. For example, consider a single-photon state $|1 \rangle$ impinging on a balanced beam splitter, the other input mode being in the vacuum state $|0 \rangle$. The two-mode output state is obtained simply by inverting transformation (\ref{BS}) and writing the input-mode annihilation
operators ${\hat a}$ and ${\hat b}$ as functions of the output-mode annihilation operators ${\hat a}'$ and ${
\hat b}'$. The input state being ${\hat a}^\dagger |0\rangle$, we can write the output state as
\begin{equation}
2^{-1/2}({\hat a}'^\dagger + {\hat b}'^\dagger) |0\rangle = (|1\rangle |0\rangle + |0\rangle |1\rangle ) / \sqrt{2},
\end{equation}
which is well known to be a path-entangled state of one photon, characterized by an entanglement entropy of 1~bit as measured by the reduced von Neumann entropy of any output mode. However, whenever we consider higher photon-number states at the input and arbitrary transmittances, it becomes much harder to find closed formulas for the entanglement entropy.

As another example illustrating the difficulty of treating a beam splitter in state space, let us consider an arbitrary input state in mode ${\hat a}$ that is coupled to a thermal field in mode ${\hat b}$. The transformation ${\hat a} \to {\hat a}'$ can be viewed as a thermal bosonic channel, which is a special case of a Gaussian phase-insensitive bosonic channel \cite{Conj04}. In order to derive the channel capacity, a crucial element is to determine the input state in mode ${\hat a}$ that results in the minimum-entropy output state in mode ${\hat a}'$.  Although it is very tempting to assume that this extremal input state is the vacuum state $|0 \rangle$, this has not been proven to date \cite{Holevo-Giovanetti}. It is linked to the Holevo-Werner conjecture, which states that Gaussian mixtures of Gaussian states achieve the capacity of such Gaussian channels \cite{Holevo-Werner}.

In this paper, we exploit majorization theory in order to study the entanglement generated by an optical beam splitter. Majorization provides a partial order relation between bipartite pure quantum states and gives a necessary and sufficient condition for the existence of a deterministic LOCC (local operation and classical communication) transformation from one state to
another \cite{Nielsen99,Nielsen01}. Here, we will show that a beam splitter obeys two classes of majorization relations, which bear some similarity to those characterizing another optical component, namely a two-mode squeezer \cite{Cerf12}. Specifically, we will prove that when feeding the input mode of a beam splitter with $k$ photons while the other input mode is in the vacuum state, the resulting two-mode output state majorizes the state corresponding to any larger photon number $k'>k$. This implies that any bipartite entanglement measure on the output modes increases with $k$ in a monotonic fashion.

Then, we examine majorization relations with respect to the coupling parameter $\theta$ (or, equivalently, the transmittance $\tau$) of the beam splitter. For a fixed photon-number input state $|k\rangle$ in one port and vacuum in the other port, we probe the existence of majorization relations between the output states corresponding to different $\theta$'s, which we call parametric majorization. We find that in a region of finite width, the output state with parameter $\theta$ majorizes the output state with $\theta' > \theta$, which implies that entanglement can only increase with the coupling between the two modes in this region. Interestingly, we also disproof parametric majorization in other regions of the parameter $\theta$, which means that the corresponding output states are then incomparable. In some cases, however, we can prove that these incomparable output states can be catalyzed \cite{Plenio99}, that is, if we supplement both states with an appropriate catalyst state, then the two product states 
become comparable (one is majorized by the other). Remarkably, the catalyst state can be as simple as a path-entangled single-photon state or a two-mode vacuum squeezed state.

The rest of this paper is organized as follows. After summarizing the basics of majorization theory in Section II, we exploit it in Section III and prove that the input state $|k\rangle |0\rangle$ yields an output state $|\Psi^{(k)}\rangle$ that majorizes the output state $|\Psi^{(k')}\rangle$ with $k'>k$.
Then, in Section IV, we proceed to investigate parametric majorization, namely the relation between the output states
$|\Psi^{(k)}(\theta)\rangle$ and $|\Psi^{(k)}(\theta+\varepsilon)\rangle$ corresponding to different transmittances.
We focus, in particular, on infinitesimal majorization relations, namely the case where $\varepsilon>0$ is infinitesimal.
We observe a default of majorization in some specific regions of parameter $\theta$,
indicating that the above states are then incomparable. We also show how majorization may be recovered by exploiting the concept of catalysis, namely by supplementing the state with an appropriate catalyst state. Finally, we close with the conclusions in Section V.

\section{Theory of majorization}

The theory of majorization gives a means to compare two probability distributions and to conclude which of the
two is more ``disordered'' or more ``random'' \cite{Muirhead03,Hardy78,Marshall79,Arnold87}.
Consider two $d$-dimensional real vectors $\textbf{p}$ and $\textbf{q}$. We say that $\textbf{p}$ is majorized by $\textbf{q}$,
symbolized by $\textbf{p}\prec \textbf{q}$, iff
\begin{eqnarray}
\sum_{i=1}^{k}p_i^{\downarrow}\leq\sum_{i=1}^{k}q_i^{\downarrow}
\label{1}
\end{eqnarray}
for $k=1,\dots,d-1$ and
\begin{eqnarray}
\sum_{i=1}^{d}p_i^{\downarrow}=\sum_{i=1}^{d}q_i^{\downarrow},
\label{2}
\end{eqnarray}
where the down-pointing arrow on $\textbf{p}$ and $\textbf{q}$ indicates that the components are sorted in non-increasing order.
Equation (\ref{2}) is automatically satisfied if $\textbf{p}$ and $\textbf{q}$ are vectors normalized to unity,
e.g., if they are probability distributions. Majorization only provides a partial order, in the sense that if $\textbf{p}$ is
\emph{not} majorized by $\textbf{q}$ (symbolized by $\textbf{p}\nprec \textbf{q}$) then this does not imply that $\textbf{p}\succ \textbf{q}$.
When both $\textbf{p} \nprec \textbf{q}$ and $\textbf{q}\nprec \textbf{p}$ hold, we say that the two vectors are \emph{incomparable}.

The definition of majorization through Eqs. (\ref{1}) and (\ref{2}) is very handy for calculation purposes, but it does not makes it clear in what sense $\textbf{p}$ is more
disordered than $\textbf{q}$. A more intuitive definition is to say that $\textbf{p}$ is majorized by $\textbf{q}$ iff there exists a set of $d-$dimensional permutation matrices
$\bm{\Pi}_n$ and a probability distribution $\{t_n\}$ such that
\begin{eqnarray}
\textbf{p}=\sum_n t_n \, \bm{\Pi}_n \cdot \textbf{q}.
\label{3}
\end{eqnarray}
Roughly speaking, the above definition says that $\textbf{p}$ is majorized by $\textbf{q}$ iff we can obtain $\textbf{p}$ by randomly permuting
the components of vector $\textbf{q}$ and afterwards taking the average over all permutations.

The equivalence between these two different definitions of majorization is implied by Theorems 1 and 2 below.
Indeed, the notion of majorization is closely related to the notion of doubly stochastic matrices. A real $d\times d$ matrix $\textbf{D}=[D_{ij}]$ is doubly stochastic if
all its entries are non-negative, and each row and each column sums up to 1. The following theorem gives the relation between majorization and doubly
stochastic matrices.\\

\textsl{Theorem 1:} $\textbf{p}\prec \textbf{q}$ iff there exists a doubly stochastic matrix $\textbf{D}$ such that $\textbf{p}=\textbf{D} \cdot \textbf{q}$.\\

The set of doubly stochastic matrices of a given dimension is convex. All extremes points of this convex set are the permutation matrices $\bm{\Pi}_n$, so any
doubly stochastic matrix can be expressed as a convex combination of permutation matrices. This is expressed in the following theorem.\\

\textsl{Theorem 2:} The $d\times d$ doubly stochastic matrices $\textbf{D}$ form a convex set whose extreme points are all the $d\times d$ permutation matrices $\bm{\Pi}_n$.\\

The convex set of $d\times d$ doubly stochastic matrices is called Birkhoff's polytope. It admits $d!$ vertices (i.e., the number of $d\times d$ permutation matrices) and its
dimension is $(d-1)^2$. Note that if we want to express a point (a doubly stochastic matrix) belonging to this $(d-1)^2$-dimensional polytope
as a convex combination of the extremal points, Caratheodory's theorem implies that we would need $(d-1)^2+1$ extremal points at most.

One naturally expects that majorization theory should be connected with various measures of ``disorder'', such as entropies. Indeed, since $\textbf{p}\prec \textbf{q}$ means that $\textbf{p}$ is more disordered that $\textbf{q}$, any measure of disorder $S:\mathbb{R}^d\rightarrow \mathbb{R}$ should satisfy
\begin{gather}
S(\textbf{p}) \ge S(\textbf{q}) \label{monotonicity}
\end{gather}
for all $\textbf{p}$ and $\textbf{q}$ such that $\textbf{p}\prec \textbf{q}$. A function $S$ obeying this property is called Schur-concave. Consider, for example, the Shannon entropy
\begin{eqnarray}
 S_1(\textbf{p})= -\sum_{i=1}^d p_i \ln  p_i
\end{eqnarray}
or the R\'{e}nyi entropy
\begin{eqnarray}
 S_{\alpha}(\textbf{p})=\frac{1}{1-\alpha}\ln \Big(\sum_{i=1}^{d} p_i^{\, \, \alpha}\Big) \label{Renyi}
\end{eqnarray}
of order $\alpha\geq 0$, $\alpha\neq 1$. (In the limit $\alpha\rightarrow 1$, the R\'{e}nyi entropy converges to the Shannon entropy.) These functions can be seen to be Schur-concave as a consequence of the following theorem \cite{Hardy78}.\\

\textsl{Theorem 3:} $\textbf{p}\prec \textbf{q}$ iff $\sum_{i=1}^d h(p_i)\geq \sum_{i=1}^d h(q_i)$ for all concave functions $h$.\\

The usefulness of majorization in quantum information theory appears first if we wish to compare two density matrices instead of probability distributions. Consider the density matrices $\rho$ and $\sigma$ of a $d-$level quantum system, and their respective vectors of eigenvalues $\bm{\lambda}(\rho)$ and $\bm{\lambda}(\sigma)$. We have the following theorem.\\

\textsl{Theorem 4:} $\bm{\lambda}(\rho)\prec \bm{\lambda}(\sigma)$ iff state $\rho$ can be obtained from state $\sigma$ by applying a mixture of unitaries.\\

The proof goes simply by noting that there is a unitary transformation that aligns the eigenbasis of $\sigma$ with that of $\rho$, and that each permutation of the eigenstates can by realized by a unitary transformation.

The connection with quantum information theory becomes even more evident in the context of comparing pure bipartite entangled states. Indeed,
majorization theory gives the means to determine whether one pure bipartite state is convertible into another one using LOCC (local operation and classical communication). Consider two $d-$level systems $A$ and $B$. Any bipartite pure states on these systems can be written in the Schmidt form
\begin{eqnarray}
|\Psi\rangle=\sum_{i=1}^d \sqrt{\lambda_i} |i\rangle_A |i\rangle_B,
\label{4}
\end{eqnarray}
where $\{|i\rangle_A\}$ and $\{|i\rangle_B\}$ are suitable orthonormal bases of the systems $A$ and $B$, respectively. The reduced density matrix of system $A$ is $\rho_A^{\Psi}\equiv \mathrm{tr}_{B}|\Psi\rangle \langle\Psi|= \sum_{i=1}^d \lambda_i|i\rangle\langle i|_A$, and similarly for $B$ (the two reduced density matrices have the same eigenvalues $\lambda_i$). We have the following theorem \cite{Nielsen99,Nielsen01}.\\

\textsl{Theorem 5:} State $|\Psi\rangle$ can be converted deterministically into state $|\Phi\rangle$ by means of LOCC iff
$\bm{\lambda}_\Psi\prec \bm{\lambda}_\Phi$, where $\bm{\lambda}_\Psi$ is the vector of eigenvalues of $\rho_A^{\Psi}\equiv \mathrm{tr}_{B}|\Psi\rangle \langle\Psi|$
and similarly for $\bm{\lambda}_\Phi$.\\

For conciseness, we will write simply $\Psi\prec \Phi$ instead of $\bm{\lambda}_\Psi\prec \bm{\lambda}_\Phi$ in what follows.
A consequence of this  theorem is that $\Psi\prec \Phi$ iff $\mu(\Psi) \ge \mu(\Phi)$ for all measures of entanglement $\mu$. A measure of entanglement, or entanglement monotone, is a non-negative function of the state which does not increase on average under LOCC and vanishes on separable states \cite{Vidal}. A common example is the entropy of entanglement, $E(\Psi)\equiv S_1(\bm{\lambda}_\Psi)$. Since, according to Theorem 5, converting $\Psi$ into $\Phi$ is possible when $\Psi\prec \Phi$, this means that $\mu(\Psi)\ge \mu(\Phi)$ must hold for all entanglement monotones. Conversely, the maximum probability of success of converting $\Psi$ into $\Phi$ by means of an LOCC protocol satisfies $P(\Psi \rightarrow \Phi)\leq \min_\mu \frac{\mu(\Psi)}{\mu(\Phi)}$, where the minimum is taken over all entanglement monotones \cite{Vidal}. Since a strategy exists where $P(\Psi \rightarrow \Phi)=1$, this implies that $\mu(\Psi) \ge \mu(\Phi)$, $\forall \mu$.

According to Theorem 5, if $\bm{\lambda}_\Psi$ and $\bm{\lambda}_\Phi$ are incomparable, then there does not exist a strategy to convert one state into the other by LOCC with probability 1. Remarkably, it has been shown that one
may nevertheless be able to accomplish such a transformation deterministically with the use of an auxiliary entangled state, an effect called catalysis \cite{Plenio99}.
If two states $|\Psi\rangle$ and $|\Phi\rangle$ have incomparable $\bm{\lambda}$ vectors, then, under certain conditions,
there exists an entangled state $|C\rangle$ that the two parties can share, called a catalyst state, such that
\begin{equation}
|\Psi\rangle \otimes |C\rangle \underset{\textrm{LOCC}}{\longrightarrow} |\Phi\rangle \otimes |C\rangle
\label{catalysis}
\end{equation}
is possible. The term ``catalysis'' is justified because the entangled state
$|C\rangle$ is recovered after the LOCC transformation.
Note that if converting $\Psi$ into $\Phi$ by catalysis is possible, then all \textit{additive} measures of entanglement must satisfy $\mu(\Psi) \ge \mu(\Phi)$. In particular,
we must have $S_\alpha(\bm{\lambda}_\Psi) \ge S_\alpha(\bm{\lambda}_\Phi) $ for all $\alpha \ge 0$.

\begin{figure}[t]
\centering
\includegraphics[scale=0.5]{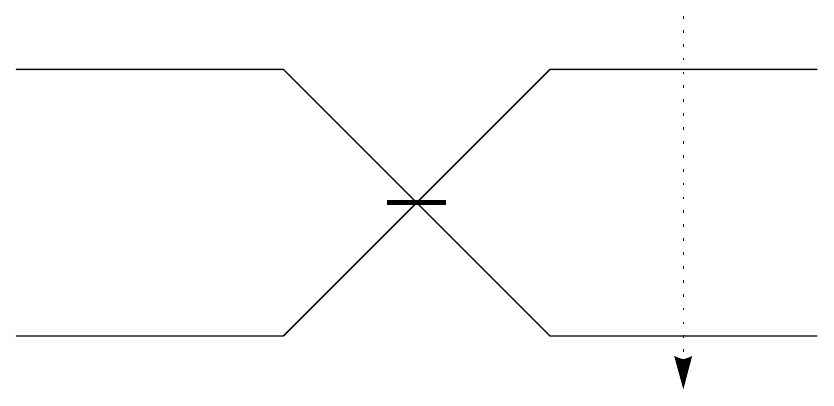}
\put(-130,8){$|0\rangle$}
\put(-130,47){$|k\rangle$}
\put(-55,26){$\theta$}
\put(-28,-5){$|\Psi^{(k)}(\theta)\rangle$}
\put(-35,-20){\phantom{1}}\\
\includegraphics[scale=0.5]{bs.pdf}
\put(-130,8){$|0\rangle$}
\put(-146,47){$|k+1\rangle$}
\put(-55,26){$\theta$}
\put(-28,-5){$|\Psi^{(k+1)}(\theta)\rangle$}
\caption{Majorization relations with respect to the input photon number. In the upper scheme, the input state is $|k,0\rangle$, while in the lower scheme it is $|k+1,0\rangle$.
The output state $|\Psi^{(k)}(\theta)\rangle$ majorizes $|\Psi^{(k+1)}(\theta)\rangle$, so that the generated entanglement $\mu(\Psi^{(k)}(\theta))$ monotonically grows with $k$ for all $\theta$.}
\end{figure}

\section{Majorization with respect to photon number}

We now prove a first class of majorization relations characteristic of a beam splitter by using the definition of majorization involving doubly stochastic matrices.
Let $|\Psi^{(k)}(\theta)\rangle$ be the output state of a beam splitter if the input state is $|k,0\rangle$, as shown in Fig.~1.
Denoting by $\mathcal{U}(\theta)$ the unitary transformation resulting from the beam splitter, we have
\begin{eqnarray}
 |\Psi^{(k)}(\theta)\rangle=\mathcal{U}(\theta) \, |k,0\rangle =\sum_{n=0}^{k}\sqrt{P_n^{(k)}(\theta)} \, |n,k-n\rangle,
 \label{5}
\end{eqnarray}
where
\begin{eqnarray}
P_n^{(k)}(\theta)=\binom{k}{n} \cos^{2n}\theta \, \sin^{2(k-n)}\theta.
\label{8}
\end{eqnarray}
The reduced density matrix of the first output mode is
\begin{eqnarray}
 \rho^{(k)}(\theta)=\sum_{n=0}^{k} P_n^{(k)}(\theta) \,  |n\rangle \langle n|,
 \label{7}
\end{eqnarray}
where $P_n^{(k)}(\theta)$ can be interpreted as the probability that $n$ photons are transmitted out of the $k$ incident photons if the transmittance of the beam splitter is $\tau=\cos^2 \theta$.

We wish to prove a majorization relation between $P_n^{(k)}(\theta)$ and $P_n^{(k+1)}(\theta)$, that is, we want to prove that
there exists a doubly stochastic matrix $\textbf{D}$ such that
\begin{eqnarray}
\textbf{P}^{(k+1)}(\theta)=\textbf{D}^{(k+1)}\cdot\textbf{P}^{(k)}(\theta),
\label{9}
\end{eqnarray}
where $\textbf{P}^{(k)}(\theta)$ is a vector having the eigenvalues $P_n^{(k)}(\theta)$ of $\rho^{(k)}(\theta)$ as components.
Note that in Eq. (\ref{9}), we wrote $\textbf{P}^{(k+1)}$ on the left-hand side and $\textbf{P}^{(k)}$ on the right-hand side, which means that $\textbf{P}^{(k+1)}$
is more disordered than $\textbf{P}^{(k)}$. Using Pascal identity for the binomial coefficients, we obtain the recurrence equation
\begin{eqnarray}
\nonumber P^{(k+1)}_{n}(\theta)&=&\binom{k+1}{n}\cos^{2n}\theta \, \sin^{2(k+1-n)}\theta\\
\nonumber&=&\Bigg(\binom{k}{n-1}+\binom{k}{n}\Bigg)\cos^{2n}\theta \, \sin^{2(k+1-n)}\theta\\
&=& P^{(k)}_{n-1}(\theta) \, \cos^{2}\theta+P^{(k)}_{n}(\theta) \, \sin^{2}\theta.
\label{10}
\end{eqnarray}
This simply expresses the fact that the probability of transmitting $n$ photons out of $k+1$ incident photons is the sum of two mutually exclusive possibilities: either $n-1$ photons (out of $k$) are transmitted and the $(k+1)$-th photon goes through, or $n$ photons (out of $k$) are transmitted and the $(k+1)$-th photon is reflected.
We can expand Eq. (\ref{10}) as
\begin{eqnarray}
\nonumber P^{(k+1)}_{0}(\theta)&=&0+\sin^{2}\theta \, P^{(k)}_{0}(\theta),\\
\nonumber P^{(k+1)}_{1}(\theta)&=&\cos^{2}\theta \, P^{(k)}_{0}(\theta)+\sin^{2}\theta \, P^{(k)}_{1}(\theta),\\
\nonumber P^{(k+1)}_{2}(\theta)&=&\cos^{2}\theta \, P^{(k)}_{1}(\theta)+\sin^{2}\theta \, P^{(k)}_{2}(\theta),\\
\nonumber &\vdots&\\
P^{(k+1)}_{k+1}(\theta)&=&\cos^{2}\theta \, P^{(k)}_{k}(\theta)+0.
\label{11}
\end{eqnarray}
This can be put in the form of Eq. (\ref{9}) by defining
\begin{eqnarray}
\textbf{P}^{(k+1)}(\theta)=\begin{pmatrix}
  P^{(k+1)}_{0}(\theta) \\
  P^{(k+1)}_{1}(\theta) \\
  P^{(k+1)}_{2}(\theta) \\
  \vdots\\
   P^{(k+1)}_{k}(\theta)\\
  P^{(k+1)}_{k+1}(\theta)
 \end{pmatrix},\ \textbf{P}^{(k)}(\theta)=
 \begin{pmatrix}
  P^{(k)}_{0}(\theta) \\
  P^{(k)}_{1}(\theta) \\
  P^{(k)}_{2}(\theta) \\
  \vdots\\
  P^{(k)}_{k}(\theta)\\
  0
 \end{pmatrix},
 \label{12}
\end{eqnarray}
where a zero entry has been inserted in the vector $\textbf{P}^{(k)}$ in order to
make $\textbf{P}^{(k)}$ and $\textbf{P}^{(k+1)}$ of equal dimension.
The doubly stochastic matrix is
\begin{eqnarray}
 \textbf{D}^{(k+1)}=\begin{pmatrix}
    \sin^2 \theta&0&0&\cdots&\cos^2 \theta\\
    \cos^2 \theta&\sin^2 \theta&0&\cdots&0\\
    0&\cos^2 \theta&\sin^2 \theta&\cdots&0\\
    \vdots&\vdots&\vdots&\ddots&\vdots\\
    0 &0 &0  &\cdots &\sin^2\theta
 \end{pmatrix},
 \label{13}
 \end{eqnarray}
where the right-most entry of the fist row has been chosen so as to fulfill the doubly-stochastic conditions (it plays no role since the last entry of  $\textbf{P}^{(k)}$ is zero).

Thus, we have proven the majorization relation
\begin{equation}
\Psi^{(k+1)}(\theta)  \prec \Psi^{(k)}(\theta),  \qquad \forall \theta,
\end{equation}
which implies that when increasing the number of the incident photons, the 2-mode output state can only be more entangled and the
corresponding 1-mode reduced states are getting more disordered. This also implies that the two parties (Alice and Bob) can convert state $ |\Psi^{(k+1)}(\theta)\rangle$ into
state $ |\Psi^{(k)}(\theta)\rangle$ by using a deterministic LOCC transformation. To achieve such a transformation,  Alice can, for example, perform a two-outcome POVM measurement with the following Kraus operators,
\begin{eqnarray}
 \mathcal{F}_1^{(k)}=\sum_{n=0}^{k}\sqrt{\frac{k+1-n}{k+1}}|n\rangle\langle n|
 \label{17}
\end{eqnarray}
and
\begin{eqnarray}
 \mathcal{F}_2^{(k)}=\sum_{n=0}^{k}\sqrt{\frac{n+1}{k+1}}|n\rangle\langle n+1|  \, ,
 \label{18}
\end{eqnarray}
satisfying $\mathcal{F}_1^{\dagger} \mathcal{F}_1+ \mathcal{F}_2^{\dagger} \mathcal{F}_2 = \openone$ for all $k$.
Then, she must communicate her outcome to Bob, who has to apply proper local unitaries.
If outcome 1 occurs, then Bob should apply the unitary
\begin{eqnarray}
 \mathcal{U}_1^{(k)}=\sum_{n=0}^{k} |n\rangle\langle n+1| +|k+1 \rangle\langle 0|
 \label{19}
\end{eqnarray}
corresponding to a cyclic shift in the space spanned by $\{|0 \rangle, \cdots |k+1 \rangle \}$.
The second term on the right-hand side of Eq.~(\ref{19}) ensures unitarity (it plays no role since Bob's reduced state is supported by $\{|1 \rangle, \cdots |k+1\rangle \}$ when outcome 1 occurs).
If outcome 2 occurs, then Bob does nothing, that is, he applies the unitary $ \mathcal{U}_2^{(k)}= \openone$.
It is easy to check that the transformation $|\Psi^{(k+1)}(\theta)\rangle\rightarrow|\Psi^{(k)}(\theta)\rangle$ takes place for both outcomes, so the LOCC transformation is indeed deterministic.

\section{Parametric majorization with respect to transmittance}
\subsection{Infinitesimal majorization}

We now examine the scenario that is summarized in Fig.~2. The input state is fixed to $|k,0\rangle$, but we change the angle $\theta$ parameterizing
the transmittance of the beam splitter by an infinitesimal amount $\varepsilon$. Note that we take $\theta\ge 0$, $\varepsilon> 0$, and $\theta+\varepsilon\leq\frac{\pi}{4}$. (For angles greater
than $\frac{\pi}{4}$, the transmittance and reflectance just interchange their roles.) An equivalent way to see this scenario is depicted in Fig.~3. Our goal is thus
to probe whether the intermediate state $|\Psi^{(k)}(\theta) \rangle$ majorizes or not the final state $|\Psi^{(k)}(\theta+\varepsilon) \rangle$.
To this end, we find it easier to use the first definition of majorization, involving the accumulations of the ordered vectors of eigenvalues
of the reduced density matrix. We will refer to these vectors as OSC (ordered Schmidt coefficients).

\begin{figure}[t]
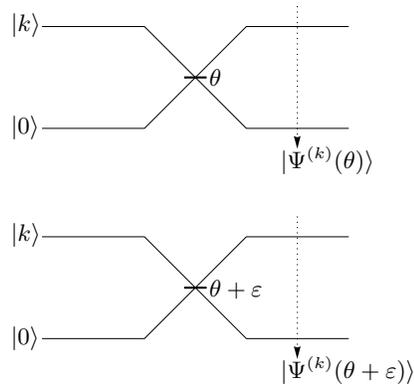

\centering
\includegraphics[scale=0.5]{bs.pdf}
\put(-130,8){$|0\rangle$}
\put(-130,47){$|k\rangle$}
\put(-55,26){$\theta$}
\put(-28,-5){$|\Psi^{(k)}(\theta)\rangle$}
\put(-35,-20){\phantom{1}}\\
\includegraphics[scale=0.5]{bs.pdf}
\put(-130,8){$|0\rangle$}
\put(-130,47){$|k\rangle$}
\put(-55,26){$\theta+\varepsilon$}
\put(-28,-5){$|\Psi^{(k)}(\theta+\varepsilon)\rangle$}
\caption{Majorization relations with respect to the coupling parameter $\theta$ (or transmittance $\tau=\cos^2\theta$).
In both schemes the input state is $|k,0\rangle$ but the angles differ by $\varepsilon$.
In some parameter region, the corresponding output states $|\Psi^{(k)}(\theta)\rangle$ and $|\Psi^{(k)}(\theta+\varepsilon)\rangle$
are proven to satisfy a majorization relation.}
\end{figure}

\begin{figure}[t]
\centering
\includegraphics[scale=0.35]{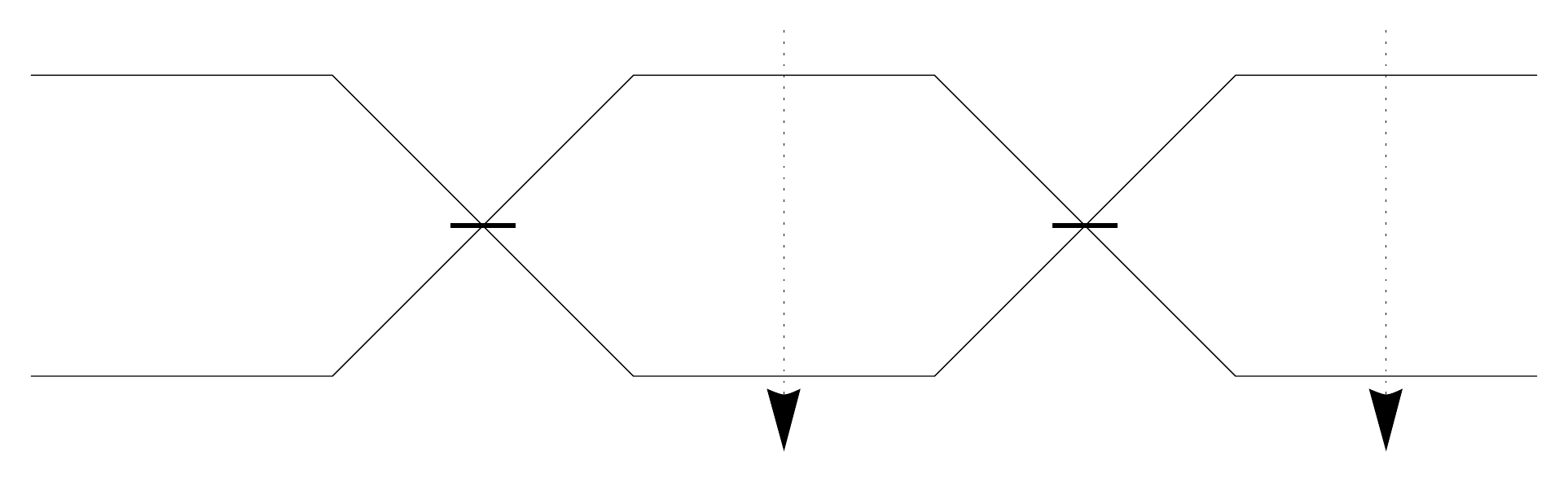}
\put(-203,11){$|0\rangle$}
\put(-203,48){$|k\rangle$}
\put(-130,29){$\theta$}
\put(-55,29){$\varepsilon$}
\put(-104,-4){$|\Psi^{(k)}(\theta)\rangle$}
\put(-29,-4){$|\Psi^{(k)}(\theta+\varepsilon)\rangle$}
\caption{Same situation as in Fig. 2, viewed as a sequence of two beam splitters with angles $\theta$ and $\varepsilon$. In some parameter region, $|\Psi^{(k)}(\theta)\rangle$ majorizes $|\Psi^{(k)}(\theta+\varepsilon)\rangle$ 
for all $k$. }
\end{figure}
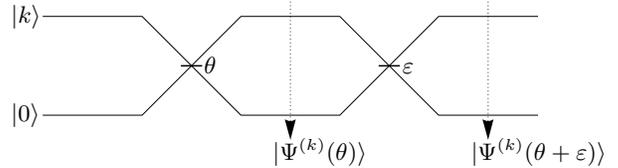

Let $\textbf{P}^{\downarrow}(\theta)$ be an OSC vector, whose components are the elements of the binomial distribution (\ref{8}).
From now on, we drop the index $k$ as it is fixed. This OSC vector will not keep the same ordering as the parameter $\theta$ varies, so we
will adopt the notation $\textbf{P}^{\downarrow r}(\theta)$, where $r=1,2,\ldots$ labels the regions of parameter $\theta$ in
which the ordering of the OSC vector remains the same. More precisely, we
have a change of ordering every time two eigenvalues $P_n(\theta)$ and $P_m(\theta)$ are equal, which occurs at
\begin{eqnarray}
  \theta=\arctan\Bigg(\frac{(k-n)!n!}{(k-m)!m!}\Bigg)^{\frac{1}{2(n-m)}}.
 \label{24}
\end{eqnarray}
Indeed, it can be shown that if $P_n(\theta)=P_m(\theta)$ for $m\neq n$, then $\frac{d P_n(\theta)}{d\theta} \neq \frac{d P_m(\theta)}{d\theta}$, i.e., the two eigenvalues cross. The cross-over points between regions are the different solutions $\theta_1<\theta_2<...$ of the above equations. We define the region $r=1$
as $\theta \in[0,\theta_1)$, the region $r=2$ as $\theta\in[\theta_1, \theta_2)$, etc.

 Our goal now is to check whether the infinitesimal majorization relation
\begin{equation}
\textbf{P}^{\downarrow r}(\theta+\varepsilon)\prec\textbf{P}^{\downarrow r}(\theta)
\end{equation}
holds or not within region $r$, taking the limit of an infinitesimal angle $\varepsilon$.
Using the definition of Eq. (\ref{1}), we have to prove
\begin{eqnarray}
 &&\sum_{n=0}^{j}P_{n}^{\downarrow r}(\theta+\varepsilon)\leq\sum_{n=0}^{j}P_{n}^{\downarrow r}(\theta) ,    \qquad j=0,\ldots k-1  \nonumber\\
 &\Leftrightarrow&\sum_{n=0}^{j}\frac{P_{n}^{\downarrow r}(\theta)}{d\theta}\leq 0 ,    \qquad j=0,\ldots k-1.
 \label{21old}
\end{eqnarray}
By defining the vector of accumulation derivatives
\begin{eqnarray}
 a_j^{\downarrow r}(\theta)=\sum_{n=0}^{j}\frac{dP_n^{\downarrow r}(\theta)}{d\theta} ,
 \label{22}
\end{eqnarray}
the infinitesimal majorization relations can be written simply as
\begin{eqnarray}
a_{j}^{\downarrow r}(\theta)\leq 0 , \qquad  j=0,\ldots k-1.
 \label{21}
\end{eqnarray}
We do not need to consider the last accumulation derivative $a_{k}^{\downarrow r}(\theta)=0$ since Eq. (\ref{2}) is necessarily verified (the OSC vectors are normalized).

The violation of at least one relation in Eqs. (\ref{21}) is sufficient to disproof majorization in region $r$. A priori, if the above majorization relations do not hold, there may nevertheless be a majorization in the opposite direction if all relations are satisfied with $\geq$ instead of $\leq$. However,  the $(k-1)$-th accumulation is the same in all regions no matter what the ordering is, and its derivative
\begin{eqnarray}
&&a_{k-1}(\theta) =-2k\sin^{2k-1}\theta \, \cos \theta
\label{30}
\end{eqnarray}
respects Eq. (\ref{21}) with a strict inequality (except in the trivial cases $k=0$ or $\theta=0$). Hence, majorization is never possible in the opposite direction.

It is easy to see that in the region $r=1$, the components of the OSC vector are
\begin{eqnarray}
 P_n^{\downarrow 1}(\theta)=\binom{k}{n}\sin^{2n}\theta\cos^{2(k-n)}\theta.
 \label{23}
\end{eqnarray}
This region extends until the second largest eigenvalue becomes equal to the largest eigenvalue and the two switch places. Beyond this cross-over point, we enter the second region $r=2$.
Now, let us prove that the infinitesimal majorization relation $\textbf{P}^{\downarrow 1}(\theta+\varepsilon)\prec \textbf{P}^{\downarrow 1}(\theta)$ always holds in region $r=1$. We have
\begin{eqnarray}
 \frac{dP_n^{\downarrow 1}(\theta)}{d\theta}=\Bigg[\frac{2n}{\tan\theta}-2(k-n)\tan\theta\Bigg]P_n^{\downarrow 1}(\theta)
\label{25}
\end{eqnarray}
and
\begin{eqnarray}
 P_n^{\downarrow 1}(\theta)=P_0^{\downarrow 1}(\theta) \binom{k}{n}\tan^{2n}\theta  ,
\label{26}
\end{eqnarray}
from which we can express the vector of accumulation derivatives as
\begin{eqnarray}
&&a_j(\theta)^{\downarrow 1}=P_0^{\downarrow 1}(\theta) \, \sum_{n=0}^{j}\Big[2n-2(k-n)\tan^2\theta\Big] \nonumber\\
 &\times&\binom{k}{n}\tan^{2n-1}\theta .
\label{28}
\end{eqnarray}
The summation in Eq. (\ref{28}) can be expressed in a closed form as
\begin{eqnarray}
&&a_j(\theta)^{\downarrow 1} = - P_0^{\downarrow 1}(\theta) \, 2(k-j)\binom{k}{j}\tan^{2j+1}\theta ,
\label{29}
\end{eqnarray}
which is non-positive for $j=0,\ldots k-1$.
Thus,  infinitesimal majorization holds within region $r=1$, which means that all states are comparable within this region,
\begin{equation}
\Psi^{(k)}(\theta+\varepsilon)  \prec \Psi^{(k)}(\theta), \qquad \forall k\ge 0,
 \end{equation}
 and all measures of entanglement increase with $\theta$.

At the cross-over point between regions $r=1$ and $r=2$, the first two components of the OSC vector switch places and the derivative of the first accumulation becomes $\frac{dP_{2}(\theta)}{d\theta}$, which is positive at this point. Therefore, majorization is violated from the left boundary of region $r=2$ until the point where this derivative ceases to be positive (and possibly beyond that point). In general, in every region that begins with a positive derivative of the first accumulation, which is equal to $\frac{dP_{n}(\theta)}{d\theta}$ for some $n$, majorization is violated at least until the point where this derivative ceases to be positive. From Eqs. (\ref{25}) and (\ref{26}), we find that the derivative
$\frac{dP_{n}(\theta)}{d\theta}$ remains positive up to the value $\theta^+_n\leq\arctan\big(\frac{n}{k-n}\big)^{\frac{1}{2n-1}}$.

In order to illustrate how parametric majorization behaves, let us exhibit three examples, corresponding respectively to a single-photon, two-photon, and three-photon state impinging on the beam splitter.

\subsection{Examples}

\noindent \textbf{Example 1:}\\
We first consider the case of a single photon ($k=1$). The OSC vector in region $r=1$ is
\begin{eqnarray}
 \textbf{P}^{\downarrow 1}(\theta)=
\begin{pmatrix}
  \cos^2\theta\\
  \sin^2\theta
 \end{pmatrix}.
 \label{31}
\end{eqnarray}
In order to find all possible regions we have to find all solutions of $\cos^2\theta=\sin^2\theta$ in $[0,\frac{\pi}{4})$. The only solution is
$\theta=\frac{\pi}{4}$, which means that there is a single region $r=1$ and, as we proved earlier, parametric majorization holds everywhere.\\

\noindent \textbf{Example 2:}\\
We now move to the case of two photons ($k=2$). The OSC vector in region $r=1$ is
\begin{eqnarray}
 \textbf{P}^{\downarrow 1}(\theta)=
\begin{pmatrix}
  \cos^4\theta\\
  2\cos^2\theta\sin^2\theta\\
  \sin^4\theta
 \end{pmatrix},
 \label{32}
\end{eqnarray}
where this ordering holds for  $[0,\arctan\frac{1}{\sqrt{2}})$. There is a second region $r=2$ corresponding to $[\arctan\frac{1}{\sqrt{2}},\frac{\pi}{4})$,
where the OSC vector is
\begin{eqnarray}
 \textbf{P}^{\downarrow 2}(\theta)=
\begin{pmatrix}
  2\cos^2\theta\sin^2\theta\\
  \cos^4\theta\\
  \sin^4\theta
 \end{pmatrix}.
 \label{33}
\end{eqnarray}
As proven in full generality, majorization holds in region $r=1$. However, in region $r=2$, the accumulation derivatives are given by
\begin{eqnarray}
 a^{\downarrow 2}_0(\theta)&=&2\cos^2\theta\sin^2\theta \nonumber\\
 a^{\downarrow 2}_1(\theta)&=&-4\cos\theta\sin^3\theta .
 \label{34}
\end{eqnarray}
The accumulation derivative $a_0(\theta)$ is positive in $[0,\frac{\pi}{4})$, so majorization does not hold in the entire region $r=2$.
This means that there ought to be measures of disorder that decrease instead of increase as a function of $\theta$ in region $r=2$. Indeed,
we observe in Fig. 4 that although the Shannon entropy increases, all other R\'{e}nyi entropies of order $\alpha>1$ exhibit a decreasing behavior somewhere within the region $r=2$. In particular,
the  R\'{e}nyi entropy of order $\alpha\rightarrow \infty$, which is the min-entropy and is directly related to the leading probability of the OSC vector, starts decreasing immediately when we enter the second region at $\theta=\arctan\frac{1}{\sqrt{2}}$.

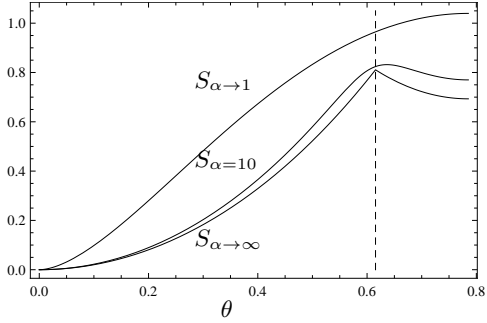
\begin{figure}[h]
\centering
\includegraphics[scale=0.6]{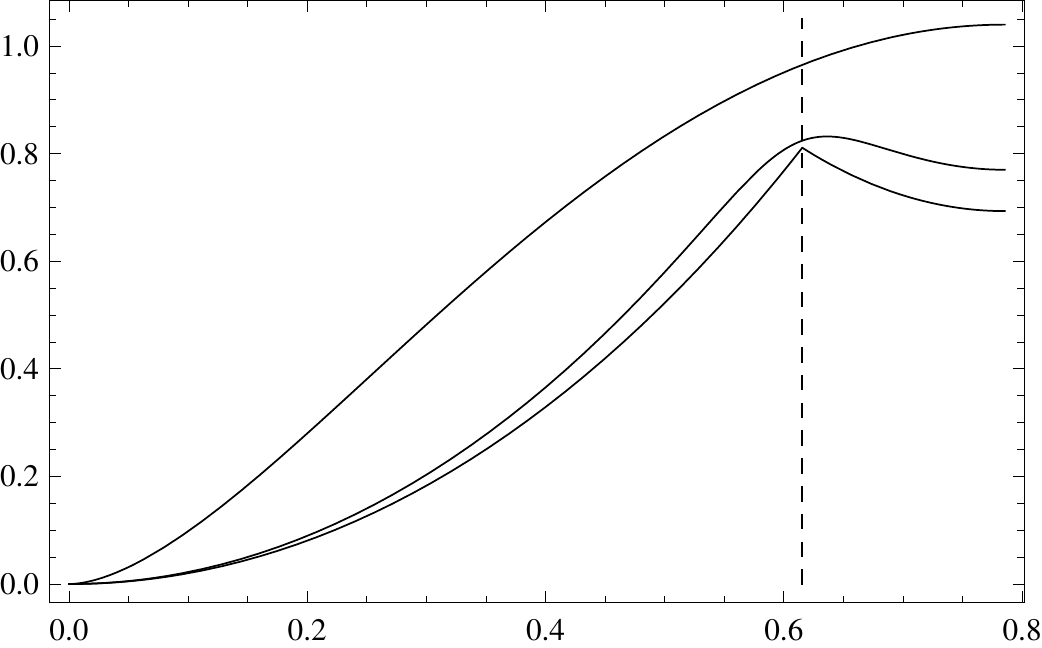}
\put(-110,80){$S_{\alpha\rightarrow1}$}
\put(-110,50){$S_{\alpha=10}$}
\put(-110,20){$S_{\alpha\rightarrow\infty}$}
\put(-100,-7){$\theta$}
\caption{Entanglement R\'{e}nyi entropies resulting from a 2-photon state impinging on a beam splitter as a function of the coupling parameter $\theta$ (related to the transmittance $\tau=\cos^2 \theta$).
The vertical dashed line denotes the boundary between parameter regions $r=1$ and $r=2$. The von Neumann entropy ($\alpha\rightarrow 1$) keeps increasing in region $r=2$,
while higher-order R\'{e}nyi entropies have a different behavior and start decreasing somewhere in region $r=2$. The min-entropy ($\alpha\rightarrow \infty$) exhibits a non-differentiable point right at the crossover point and decreases throughout the entire region $r=2$, reflecting the default of majorization.}
\end{figure}

\noindent \textbf{Example 3:}\\
As a last example, we consider the case of three photons ($k=3$). We have two cross-over angles,
\begin{eqnarray}
&&\theta_1=\arctan\frac{1}{\sqrt{3}}\nonumber\\
&&\theta_2=\arctan\frac{1}{\sqrt[4]{3}},
\label{36}
\end{eqnarray}
which define three regions of different orderings in $\big[0,\frac{\pi}{4}\big)$
and three corresponding OSC vectors,
\begin{eqnarray}
 \textbf{P}^{\downarrow 1}(\theta)&=&
\begin{pmatrix}
  \cos^6\theta\\
  3\cos^4\theta\sin^2\theta\\
  3\cos^2\theta\sin^4\theta\\
  \sin^6\theta
 \end{pmatrix} \nonumber\\
 \textbf{P}^{\downarrow 2}(\theta)&=&
 \begin{pmatrix}
  3\cos^4\theta\sin^2\theta\\
  \cos^6\theta\\
  3\cos^2\theta\sin^4\theta\\
  \sin^6\theta
 \end{pmatrix} \nonumber\\
 \textbf{P}^{\downarrow 3}(\theta)&=&
 \begin{pmatrix}
  3\cos^4\theta\sin^2\theta \\
  3\cos^2\theta\sin^4\theta\\
  \cos^6\theta\\
  \sin^6\theta
 \end{pmatrix}.
 \label{37}
\end{eqnarray}
In region $r=1$, it is easy to confirm that majorization holds, as it should. In regions $r=2$ and $r=3$,
the accumulation derivatives are given, respectively, by
\begin{eqnarray}
 a^{\downarrow 2}_0(\theta)&=&3 \cos^3\theta (-1 + 3 \cos 2 \theta) \sin \theta \nonumber\\
 a^{\downarrow 2}_1(\theta)&=&-\frac{3}{2} \sin^3 2 \theta \nonumber\\
 a^{\downarrow 2}_2(\theta)&=&-6 \cos\theta \sin^5\theta
 \label{38}
\end{eqnarray}
and
\begin{eqnarray}
 a^{\downarrow 3}_0(\theta)&=&3\cos^3\theta (-1 + 3 \cos 2 \theta) \sin \theta \nonumber\\
 a^{\downarrow 3}_1(\theta)&=&\frac{3}{2} \sin 4 \theta \nonumber\\
 a^{\downarrow 3}_2(\theta)&=&-6 \cos\theta \sin^5 \theta .
 \label{39}
\end{eqnarray}
The quantity $a^{\downarrow 2}_0(\theta)$ is positive in the interval $[0,\arctan\frac{1}{\sqrt{2}})$, which means that we are sure that
there is no majorization in the interval $[\arctan\frac{1}{\sqrt{3}},\arctan\frac{1}{\sqrt{2}})$, i.e., from the left boundary of region $r=2$ up to where $a^{\downarrow 2}_0(\theta)$
remains positive. In region $r=3$, $a^{\downarrow 3}_2(\theta)$ is always positive,
while the other accumulation derivatives are negative within this region. Hence, for $r=3$, the states are always incomparable.
This last violation of majorization is, however, not visible with the R\'{e}nyi entropies. In Fig. 5, we display the evolution of entropies across the three regions.\\

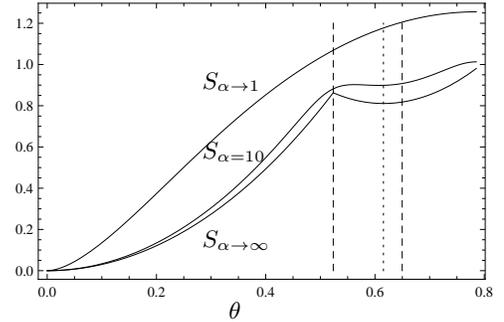
\begin{figure}[H]
\centering
\includegraphics[scale=0.6]{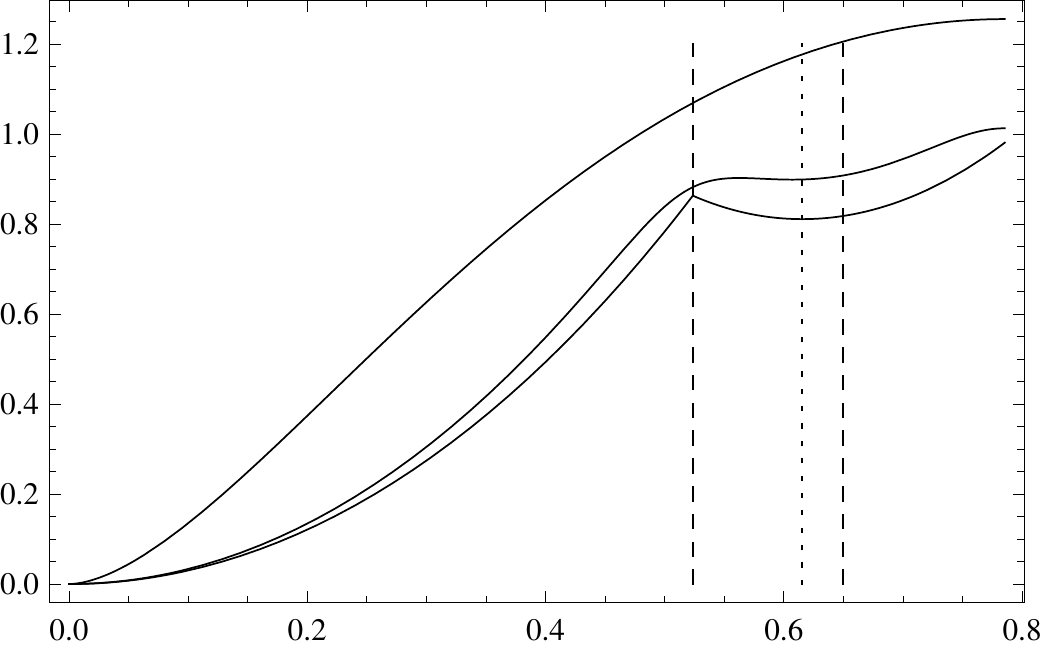}
\put(-110,80){$S_{\alpha\rightarrow1}$}
\put(-110,53.5){$S_{\alpha=10}$}
\put(-110,20){$S_{\alpha\rightarrow\infty}$}
\put(-100,-7){$\theta$}
\caption{Entanglement R\'{e}nyi entropies resulting from a 3-photon state impinging on a beam splitter as a function of $\theta$. The two vertical dashed lines at $\theta_1$ and $\theta_2$ separate the three regions $r=1,2,3$, while the dotted line corresponds to the local minimum of the min-entropy. Majorization is violated in the region $r=2$ from the left boundary of this region at $\theta_2$ up to the local minimum of the min-entropy. The majorization violation throughout the entire region $r=3$ is not manifested by the behavior of the R\'{e}nyi entropies.}
\end{figure}

\subsection{Catalysis}

We have shown that one can always expect a default of majorization when the leading term in the probability vector changes, and this majorization default prevails at least up
to the point where the first accumulation derivative ceases to be positive, or, equivalently, until we reach the local minimum of the min-entropy within this region.
Beyond the case where the first two components of the probability vector switch places, it appears difficult to provide general rules for predicting the existence or absence of majorization for an arbitrary $k$, and one has to treat the problem on a case-by-case basis. The situation also becomes more complicated if we take a non-infinitesimal angle $\varepsilon$ such that
the pair of angles $\theta$ and $\theta+\varepsilon$ belong to different regions.

It is natural to ask whether the incomparable states that occur when we change the parameter $\theta$ can nevertheless be made comparable through catalysis. We will show that this is indeed possible in certain cases, and will provide an example for this. Note that not all incomparable states can be catalyzed: some necessary conditions have to be respected \cite{Plenio99,Daftuar01,Zhou00,Sun05}. To solve the problem of whether catalysis is possible or not in generality is difficult due to the fact that one has to reorder the vector resulting from the tensor product of the state to be catalyzed and the catalyst state.

\begin{figure}[t]
\centering
\includegraphics[scale=0.6]{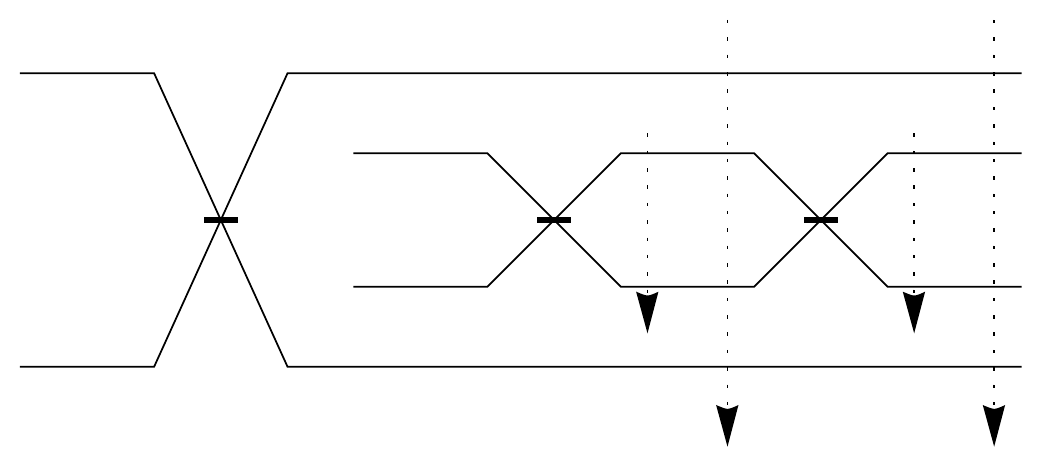}
\put(-188,66){$|1\rangle$}
\put(-188,15){$|0\rangle$}
\put(-139,40.5){{\tiny 0.7}}
\put(-131,52){$|3\rangle$}
\put(-131,29){$|0\rangle$}
\put(-81,40.5){{\tiny 0.62}}
\put(-34,40.5){{\tiny 0.1}}
\put(-115,-2){{\tiny $|\Psi^{(3)}(0.62)\rangle\otimes|C(0.7)\rangle$}}
\put(-39,-2){{\tiny $|\Psi^{(3)}(0.72)\rangle\otimes|C(0.7)\rangle$}}
\put(-93,19){{\tiny $|\Psi^{(3)}(0.62)\rangle$}}
\put(-46,19){{\tiny $|\Psi^{(3)}(0.72)\rangle$}}
\caption{Schematic of the catalyzed conversion between the incomparable states resulting from a 3-photon state impinging on a beam splitter with angles
$\theta=0.62$ and $\theta+\omega=0.72$. The catalyst is the entangled state obtained from a beam splitter with angle $\theta=0.7$ and one single-photon input state.}
\end{figure}
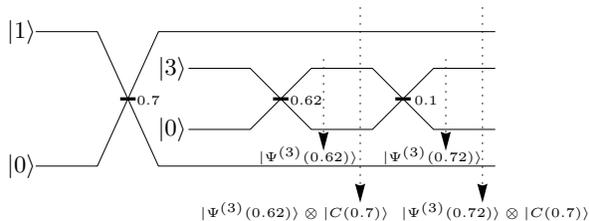

Consider two angles $\theta$ and $\theta+\omega$ that give OSC vectors $\textbf{P}^{\downarrow r}(\theta)$ and $\textbf{P}^{\downarrow r'}(\theta+\omega)$
which are incomparable. These angles may be within different ordering regions, which is the case in the following example where $r=2$ and $r'=3$. We take
$k=3$, $\theta=0.62$, and $\omega=0.10$, which gives
\begin{eqnarray}
\textbf{P}^{\downarrow 2}(0.62)=
\begin{pmatrix}
0.44439\\0.290641\\0.226491\\0.0384782
\end{pmatrix}
\label{40}
\end{eqnarray}
and
\begin{eqnarray}
\textbf{P}^{\downarrow 3}(0.72)=
\begin{pmatrix}
0.416698\\0.320544\\0.180565\\0.0821927
\end{pmatrix} 
\label{41}
\end{eqnarray}
such that $\textbf{P}^{\downarrow 3}(0.72)\nprec \textbf{P}^{\downarrow 2}(0.62)$.
The path-entangled single-photon state $\cos \theta |1\rangle |0\rangle + \sin \theta |0\rangle |1\rangle$
resulting from a single photon impinging on a beam splitter with angle $\theta = 0.7$ is sufficient to serve as a catalyst
for these two states. It corresponds to the binary probability vector
\begin{eqnarray}
\textbf{C}(0.7)=
\begin{pmatrix}
0.584984\\0.415016
\end{pmatrix}  ,
\label{42}
\end{eqnarray}
and one can easily verify from Eqs. (\ref{40}), (\ref{41}), and (\ref{42}),  that
$\textbf{P}^{\downarrow 3}(0.72)\otimes\textbf{C}(0.7)\prec \textbf{P}^{\downarrow 2}(0.62)\otimes\textbf{C}(0.7)$, implying that the catalyzed conversion is possible.  This is
summarized in Fig. 6. We could also use as a catalyst a two-mode squeezed vacuum state $\propto \sum_{n=0}^\infty \tanh^n r \, |n\rangle |n\rangle$ with squeezing parameter $r=1.38$.
Several other numerical examples can be found and some of them, like the ones provided above, are experimentally accessible.

Note that, due to the additivity of the R\'{e}nyi entropies, one should look for catalyzable incomparable states only in regions where all of the R\'{e}niy entropies increase (see also Refs.~\cite{Nechita08,Duan05}). However, in the case at hand, we have got numerical evidence that the sole behavior of the min-entropy seems to give a necessary and sufficient condition for the existence of catalysis. The latter property will be examined in a forthcoming work.

\section{Conclusion}

We have found several classes of majorization relations characterizing a beam splitter, or more generally the linear coupling between a pair of bosonic modes. More formally, we have proven that the passive Bogoliubov transformation of Eq. (\ref{BS}) fulfills some majorization relations, which enable comparing the output states corresponding to various input photon numbers $k$ as well as various coupling parameters $\theta$ (or transmittances $\tau=\cos ^2 \theta$). Interestingly, this behavior is reminiscent of the majorization relations that have recently been shown to prevail with an active Bogoliubov
transformation (a two-mode squeezer or parametric amplifier) \cite{Cerf12}. Note that in contrast to Ref. \cite{Cerf12}, the present analysis has no implications on the Holevo-Werner conjecture. Indeed, we consider situations where one of the input modes is in the vacuum state, corresponding, in the language of quantum channels, to the case of a pure lossy channel. For that specific channel, the conjecture is well known to be true \cite{lossy} since the vacuum input state results in another vacuum state at the output of the channel (the vacuum is the extremal input state since the corresponding output entropy is minimum---it is zero).

We have shown that for any value of the transmittance parameter $\theta$, the output states resulting from injecting Fock states $|k\rangle$ in one port of the beam splitter and vacuum $|0\rangle$ in the other port obey a chain of majorization relations $\Psi^{(k+1)}(\theta)  \prec \Psi^{(k)}(\theta)$,  for all $k\ge 0$. As a consequence, the output states can only be more entangled when increasing the number of incident photons, and we have found an explicit deterministic LOCC transformation that maps $\Psi^{(k+1)}(\theta)$ onto $\Psi^{(k)}(\theta)$.

In contrast, we have found that the situation is more complicated when varying the parameter $\theta$ and keeping $k$ constant. In that case, we have shown that there exists a first region in the space of parameter $\theta$ where a parametric infinitesimal majoriation relation holds, by taking the limit of an infinitesimal angle $\varepsilon>0$ for any $k\ge 0$, namely
$\Psi^{(k)}(\theta+\varepsilon)  \prec \Psi^{(k)}(\theta)$. This implies a monotonic increase of the entanglement of the output states when decreasing the transmittance and moving towards a balanced beam splitter. However, beyond some value of the parameter $\theta$, we have shown the existence of a default of majorization, which occurs because the ordering of the OSC vectors changes in such a way that the leading probability is replaced by another one. This majorization default holds from the left boundary of this new ordering region at least up to the local minimum of the min-entropy. Moreover, by examining specific examples, we have shown that one may find more violations of majorization for non-infinitesimal angles $\varepsilon$ within the same ordering region or between different ordering regions.

Finally, we have provided an example of two incomparable states, resulting from different values of $\theta$, whose conversion can nevertheless be catalyzed with the help of an experimentally accessible state, such as a single-photon path-entangled state or a two-mode squeezed vacuum state.
Catalysis schemes like the one in Fig. 6 may potentially be used for authentication protocols based on entanglement-assisted LOCC \cite{Barnum99,Shack00}.
Further investigations should also include a more general solution to the catalysis process in the parameter-varying case,
the analysis of majorization relations in more complicated optical circuits in the spirit of \cite{Latorre02}, or more ambitiously the study of phase transitions and critical phenomena in a field-theoretical approach \cite{Scholl05} under the prism of parametric majorization, where the parameter could be the temperature of a thermal field \cite{Gagatsos12}.

\begin{acknowledgments}
We thank R. Garc\'{i}a-Patr\'{o}n for useful discussions. This work was supported by the F.R.S.-FNRS under the Eranet project HIPERCOM.
C.N.G. acknowledges financial support from Wallonia-Brussels International via the excellence grants
programme. O.O. acknowledges the support of the European Commission under the Marie Curie Intra-European
Fellowship Programme (PIEF-GA-2010-273119).
\end{acknowledgments}

\bibliographystyle{apsrev}

\end{document}